\documentstyle[11pt,aaspp4]{article}

\begin{document}
\title{ Spiral Structure Based Limits on the Disk Mass \\
of the Low Surface Brightness Galaxies UGC~6614 and F568-6}

\author{
A.~C.\ Quillen\altaffilmark{1}$^,$\altaffilmark{2} \&
T.~E.\ Pickering\altaffilmark{1}$^,$\altaffilmark{3}
}
\altaffiltext{1}{The University of Arizona, Steward Observatory, Tucson, AZ 85721}
\altaffiltext{2}{E-mail: aquillen@as.arizona.edu}
\altaffiltext{3}{E-mail: tim@as.arizona.edu}

\def\spose#1{\hbox to 0pt{#1\hss}}
\def\lta{\mathrel{\spose{\lower 3pt\hbox{$\mathchar"218$}}
     \raise 2.0pt\hbox{$\mathchar"13C$}}}
\def\gta{\mathrel{\spose{\lower 3pt\hbox{$\mathchar"218$}}
     \raise 2.0pt\hbox{$\mathchar"13E$}}}

\begin{abstract}
The spiral structure of the low surface brightness galaxies F568-6 (Malin 2) 
and UGC~6614 is large scale, with arms that wrap more than half a
revolution, and extend out to 50 and 80 kpc in UGC~6614 and F568-6
respectively.  The density contrasts observed in the \ion{H}{1} maps
are high, with arm/interarm contrasts of $\sim$2:1, whereas the
velocity perturbations due to spiral structure are low, in the range
10--20 km/s and 10--30 km/s in UGC~6614 and F568-6 respectively.

Upper limits for the disk mass-to-light ratios are estimated by
considering the minimum velocity perturbations in the \ion{H}{1}
velocity field that should result from the spiral structure observed
in the $R$ band images.  
The weak observed response in the
$\phi$ velocity component limits the mass-to-light ratios of the disk
inside a scale length
to $M/L \lta 3$ and 6 for UGC~6614 for F568-6
respectively (in solar units) based upon azimuthal variations observed
in the $R$ band images.  These limits are sufficiently strong to
require a significant dark matter component even in the central
regions of these galaxies. 
Our limits furthermore imply that this dark matter component
cannot be in the form of a cold disk since a cold disk would
necessarily be involved in the spiral structure.
However, a more massive disk could be consistent with the observations 
because of a non-linear gas response or if the gas 
is driven by bar-like distortions instead of spiral structure.

To produce the large observed arm/interarm \ion{H}{1} density
variations it is likely that the spiral arm potential perturbation is
sufficiently strong to produce shocks in the gas.  For a forcing that
is greater than $2\%$ of the axisymmetric force, $M/L \gta 1$ is
required in both galaxies in the outer regions.  This is equivalent to
a disk surface density between $r =$ 60--120\arcsec\ in UGC~6614 of
2.6--1.0 $M_\odot/{\rm pc}^2$ and between $r =$ 40--90\arcsec\ in
F568-6 of 6.6--1.0 $M_\odot/{\rm pc}^2$ assuming that the amplitude of
the variations in the disk mass is the same as that observed in the $R$
band.  These lower limits imply that the stellar surface density is at least
of the same order as the gas surface density.  This is consistent with
the large scale morphology of the spiral structure, and the stability
of the gas disk, both which suggest that a moderate stellar component
is required to produce the observed spiral structure.

\end{abstract}


\section {Introduction}

The observed rotation curve shapes of both giant, so-called
Malin-type, low surface brightness galaxies (e.g. \cite{deb96a},
\cite{pic97}) and dwarf low surface brightness galaxies
(e.g. \cite{deb96a}, \cite{vz97}) do not match the rotation curves
shapes inferred from their light distributions assuming constant
mass-to-light ratios. In both cases, the observed rotation curves rise
more slowly and continue rising out to radii where the rotation curves
inferred from the light distributions are declining. In 
the giant low surface brightness galaxies, very high disk
mass-to-light ratios of about 20--30 (in $R$ band) would be required to match 
the high observed rotation speeds (\cite{pic97}).  As a result, to fit the
rotation curves of these galaxies a substantial dark matter component
is required, even at small radii, and they are said to be dark matter
dominated.  This is in contrast to normal high surface brightness
galaxies where maximal disk solutions with lower mass-to-light
ratios yield good fits to the rotation curves (e.g. \cite{ken87a} \&
\cite{ken87b}) in the central few disk scale lengths.  Comparison of
low and high surface brightness galaxies with equal total luminosity
suggest that low surface brightness galaxies have low mass surface
density disks (\cite{deb96b}).  This is consistent with the fact that
these galaxies appear at near normal locations on the Tully Fisher relation
(\cite{zwa95}).

In spite of their low mass surface densities, some low surface
brightness disk galaxies do show spiral structure including, of
course, the two cases discussed here as well as all of the giant low
surface brightness galaxies presented in \cite{spray95} and some of
the dwarfs in the samples of \cite{vz97} and \cite{deb96b} (e.g.\ UGC
11820, UGC~5716, and F568-1).  Since spiral structure requires
non-axisymmetric mass perturbations in the disk of the galaxy itself,
its properties can be used to limit the {\it mass} of the {\it disk}
involved in the spiral density waves.  If velocity perturbations
caused by the spiral structure are small, a limit can be placed on the
total mass in the spiral structure, yielding an upper limit on the
mass-to-light ratio of the luminous stellar disk.  
If evidence for strong spiral
shocks is seen in the gas response, then a sufficiently strong spiral
gravitational force is required to produce this gas response.  This
critical forcing yields a lower limit for the mass of the disk.

Indeed the concept of using spiral arm patterns to limit the disk
mass-to-light ratio was considered by \cite{vis80} who compared the
predictions of spiral density wave theory to observations of M81.  He
stated ``The second test is whether the amplitude of the wave as
measured'' (by photometry) ``converted into the amplitude of the
spiral arm potential perturbation, is consistent with the observed
amplitude of the density and velocity perturbations of the gas.''  He
treated the forcing spiral wave amplitude as a free parameter and
found that large amplitudes produced velocity residuals too large to
be consistent with the observations, whereas weak forcing amplitudes
did not produce shocks leading to the large density contrasts which
are observed in \ion{H}{1} emission in M81.  Using a different
approach, \cite{ath87} placed limits on the mass-to-light ratios of
some normal disk galaxies by assuming that the disks should amplify
$m=2$ or bisymmetric spiral modes but inhibit asymmetric $m=1$
modes.

Low surface brightness galaxies are an important setting to place
limits on the mass in the disk for two reasons: 1) They have low
density disks and 2) They appear to be dark matter dominated.
Normal high surface brightness galaxies with strong spiral structure
have bright disks which are adequately massive to produce a strong
gravitational spiral force even when the spiral structure is
relatively weak.  However, in the low surface brightness galaxies the
disk surface brightness is 1--3 magnitudes fainter even though
rotational velocities are nearly equivalent so it is quite surprising
that spiral structure exists in any of these galaxies.  Many normal
galaxies can be well fit by maximal disk or near maximal disk rotation
curve models (e.g. \cite{ken87a}, \cite{ken87b}) and models of spiral
gas response require gravitational forcing which is consistent with the 
amplitude variations observed from images of these disks (e.g. \cite{low94}).
By placing limits on the mass density of the disk from the existence
of spiral structure we can test the degree to which these galaxies are
dark matter dominated and the possibility that these galaxies might
have a very massive disk component.

In this paper we use the spiral structure observed in optical and
\ion{H}{1} images of the low surface brightness galaxies UGC~6614 and
F568-6 (Malin 2) to place both upper and lower limits on the mass-to-light ratio
of the optical disk.  In \S 2 we review the \ion{H}{1} and $R$ band
images which show spiral structure in these two galaxies.  In \S 3 we
place upper limits on the mass-to-light ratio of this spiral structure
based on the low level of spiral arm induced velocity perturbations.
In \S 4 we consider the mass density in spiral structure required to
produce shocks in the gas that would be consistent with the large
density contrasts observed in the \ion{H}{1} column density maps.  A
discussion follows in \S 5.

\section{Spiral Structure in UGC~6614 and F568-6}

UGC~6614 and F568-6 were observed in $R$ band and in \ion{H}{1} by
\cite{pic97} to investigate their neutral hydrogen and kinematic
properties.  These data are displayed as overlays in Figures 1 and 2
and are described in detail in \cite{pic97}.   Table 1 lists some
observation parameters and basic properties from 
\cite{pic97}.  UGC~6614 and F568-6 are
both low surface brightness galaxies with central surface brightnesses
several magnitudes fainter than sky level ($\mu_{R}(0) = 22.9$ and
22.1 for UGC~6614 and F568-6 respectively). Their disk exponential
scale lengths are large (14 and 18 kpc for UGC~6614 and F568-6
respectively) and they both contain copious amounts of \ion{H}{1}
($2.5$ and $3.6 \times 10^{10} M_\odot$ for the two galaxies
respectively, \cite{pic97}). We assume here the distances of $D=85$ and
$184$ Mpc to UGC~6614 and F568-6 respectively (following \cite{pic97};
derived from a Hubble constant of $75$ km s$^{-1}$ Mpc$^{-1}$).

Both galaxies show clear evidence of spiral structure (see Figures 1
and 2).  Spiral arms are visible in the optical images, the \ion{H}{1}
column density maps, and as kinks in the velocity field at the level of
10--30 km/s.  This spiral arm structure is coincident in the $R$ band
and \ion{H}{1} images as well as in the velocity field.  The spiral
structure for the two galaxies is large scale extending to a radius of
more than 50 and 90 kpc in UGC~6614 and F568-6 respectively.  This is
in contrast to normal galaxies such as M81 where the entire spiral
arm structure lies within the optical disk or within $\sim 20$ kpc.
We note however that in terms of disk scale lengths the
extent of spiral structure in UGC~6614 and  F568-6 is not so large,
extending only to 3.6 and 5 scale lengths respectively, which
is small compared to 8 scale lengths in M81 
(the disk exponential scale length in M81 $\sim 2.5$ kpc, \cite{ken87b}).

\subsection{Spiral Arm Morphology}

For both galaxies the morphology of the spiral structure consists of
coherent spiral arms each of which wrap around the galaxy more than
half a revolution.  In this respect the spiral structure does not
resemble that of flocculent late-type galaxies where pieces of arms
exist only locally.  Both spiral structures are strongly asymmetric,
particularly in their inner regions.  The outer arms in UGC~6614,
however, are close to being a bi-symmetric or two arm spiral pattern
which is centered about a position located to the east of the nucleus.
In flocculent galaxies, the spiral structure is thought to be
primarily propagating in the gas disk without strong coupling to the
stellar disk (e.g. \cite{bra93}).  Also, simulations have shown that
when the disk mass is low, the spiral structure is more likely
to be flocculent (\cite{sel84}, \cite{car85}).  The more coherent
nature of the spiral structure observed in these low surface
brightness galaxies suggests that the stellar disk is coupled to the
gas disk and actively involved in the spiral structure.

The \ion{H}{1} column density map shows large arm/interarm contrasts
of $\sim$ 2:1 and velocity variations seen as kinks in the velocity
field detectable at the level of 10--20 and 10--30 km/s in UGC~6614
and F568-6 respectively.  We note that these values for the
arm/interarm contrast and velocity residuals are similar to those
observed in M81 which has radial velocity perturbations of $\sim 10$
km/s observable in \ion{H}{1} (\cite{vis80}).  High surface brightness
galaxies can also have significantly stronger velocity perturbations.
For example, M51 has radial velocity perturbations of 60--90 km/s seen
in the CO velocity field (\cite{vog88}), and UGC~2885 has
perturbations of 50--70 km/s observed in H$\alpha$ (\cite{can93a}).

Along the kinematic minor axis of the galaxy velocity residuals are
expected to be radial.  The pattern of the velocity residual pattern
depends on whether the spiral arm pattern lies within the corotation
radius (\cite{can93}).  Kinks in the velocity field alternate sign
only a few times as a function of radius along the minor axes of the
two galaxies.  This residual pattern which is consistent with a single
residual alternating sign pair associated with each arm is similar to
the velocity residual field of M81 and suggests that the entire spiral
pattern lies within the corotation resonance (\cite{can93}).


Low surface brightness galaxies can have gas densities (\cite{vdh93}, \cite{mcg92})  
which fall below the threshold $\Sigma_{crit}$ which \cite{ken89} found was
required for massive star formation in normal spiral galaxies.
As proposed by \cite{ken89} this critical density is directly related 
to the local stability 
of the disk where the Toomre stability parameter $Q$, defined as
\begin {equation}
Q \equiv {\kappa \sigma \over 3.36 G \Sigma  }
\end{equation}
(see \cite{B+T}) can be written in terms of the critical gas density as
\begin {equation}
Q = {\Sigma_{crit} \over \Sigma \alpha}.
\end{equation}
Here $\Sigma$ is the local gas surface density, $\kappa$ is the epicyclic frequency, 
$\sigma$ is the velocity dispersion, and
$\alpha$ was emperically determined by \cite{ken89} to be $\sim 0.7$.
When $Q > 2$ amplification 
processes such as the swing amplifier are inefficient and the disk is unresponsive 
to tidal perturbations which could excite spiral density waves in a more 
unstable disk (see \cite{B+T} and references theirin).
It is therefore unlikely for a disk with $Q>2$ to show spiral structure.
This implies that a disk with gas density below the critical gas density
of \cite{ken89}
is also unlikely to show spiral structure assuming there is no other
disk mass component.  (Note that $Q\sim 1.4 $ for $\Sigma_{crit} /\Sigma = 1$.)
Consequently if a gas disk is well below the critical gas density
and yet shows spiral structure, a natural explanation is that there
is another massive component in the disk
(see \cite{jog84} for instability in a two fluid disk).

Low surface brightness galaxies can show spiral structure despite the fact that
their gas densities fall below the critical density
(\cite{vdh93}, \cite{deb96b}).   Unfortunately the \ion{H}{1} data of \cite{pic97} is not
of sufficiently high velocity resolution to measure the gas velocity dispersion 
$\sigma$ required to calculate the $\Sigma_{crit}$ predicted by \cite{ken89}.
However, \cite{pic97} found that even if $\sigma$ had the low value
of 6km/s, UGC~6614 had gas densities below the critical density  
for $r <50''$, $r>120''$ 
and that Malin 2 had gas densities below this threshold everywhere.  
This suggests that a gas plus stellar disk
may be required for these disks to be sufficiently unstable to support the
observed spiral structure.

\subsection{Spiral Arm Amplitudes}

To estimate the amplitude of the spiral arm structure observed in
\ion{H}{1} and in the $R$ band images, we must first correct for the
inclination of the galaxy.  This is relatively straightforward in the
case of UGC~6614 since the $R$ band isophote, \ion{H}{1} isophote and
\ion{H}{1} velocity field derived position and inclination angles all
agree (PA = $116^\circ$, $i=35^\circ\pm 3 ^\circ$, \cite{pic97}).  For
F568-6 the $R$ band isophote and \ion{H}{1} velocity field derived
position and inclination angles agree (PA = $75 ^\circ$,
$i=38^\circ\pm 3 ^\circ$), though the \ion{H}{1} distribution is more
asymmetric than the $R$ band light (\cite{pic97}).  We have adopted
these orientations to correct for the galaxy inclinations.

In Figures 3 and 4 we show azimuthal profiles in the inclination
corrected $R$ band images and \ion{H}{1} maps for the two galaxies.
The amplitudes as a function of radius of the $m=1$ and $m=2$
azimuthal Fourier components expressed as a percentage of the
azimuthal average for the two galaxies are shown in Figure 5 for the
$R$ band images and in Figure 6 for the \ion{H}{1} images.  UGC~6614
has an oval distortion in its central region ($r < 40''$) which is
particularly noticeable after correction for inclination.  (By oval
distortion we mean an elongation in the isophotes that does not vary
in position angle over a range of radius.)  This oval distortion is
seen as a large azimuthal density variation (evident as high $m=2$
components in Figure 5) which is not an actual intensity variation
in the spiral arms.  In the outer regions of both galaxies the faint
spiral arms are far brighter than the underlying disk which is only
marginally detected.  Although peaks in the \ion{H}{1} and $R$ band
surface brightness are correlated (see Figures 1 and 2) they do not
correspond to maximum densities or surface brightnesses in the
azimuthal cuts except at large radii in UGC~6614 (see Figures 3 and
4).  Better correlation might be observed with higher angular
resolution \ion{H}{1} observations.

Since the amplitudes shown in Figures 3 and 4 depend on the assumed
galaxy orientations we recomputed them for moderate variations in the
inclination angles.  For a $5^\circ$ lower inclination of $30^\circ$
in UGC~6614, the change in amplitudes were largest in the region of
the oval distortion ($r<40''$) and were $10-15\%$ smaller than at an
inclination of $35^\circ$.  Variations in the amplitudes elsewhere in
UGC~6614 and in F568-6 for a corresponding difference in inclination
were smaller, $\lta 5\%$.  These amplitude uncertaintites are not
sufficiently large to significantly change the limits for the
mass-to-light ratio we estimate below.

\section{Placing an Upper Limit on the Mass-to-Light Ratio}

Spiral density waves are a resonant wave phenomenon.  Because
resonances can exist a small gravitational perturbation can give a
large response, however the opposite is not true.  Given a particular
spiral gravitational perturbation there is a minimum possible
response.  This makes it possible to place an upper limit upon the
strength of the gravitational perturbation if a measure of the
response, such as kinks in the velocity field, are small.
To place an upper limit on the mass-to-light ratio of the spiral
structure, we consider that the mass involved in the spiral density
wave is insufficient to drive a strong response.  We therefore estimate
the smallest possible gas response to a 
non-axisymmetric or spiral gravitational perturbation.  For such a 
perturbation, an expansion to first order should give an
appropriate description for the gas flow in 
regions not directly affected by resonances.  In this section we
follow the notation and derivation given in chapter 6 of \cite{B+T},

We assume that there is a non-axisymmetric perturbation to the
gravitational potential in the plane of the galaxy of the form
\begin {equation}
\Phi_1( r,\phi) = Re[\Phi_{a}(r) \exp{i(m\phi - \omega t)}]
\end {equation}
where the pattern speed of the perturbation is $\Omega_p \equiv
\omega/m$.  We consider perturbations in the gas velocity field in response
to this perturbation.
The response in velocity to a tightly wound spiral density wave to
first order for the $\phi$ (non-radial) component is of the same form
as that of the potential and
in the tight winding or WKB approximation is (\cite{B+T}, equation
6-37)
\begin {equation}
v_{\phi a}(r) = {-2Bik(\Phi_a + h_a) \over 
\kappa^2 - (m \Omega - \omega)^2} 
\end {equation}
where $\kappa$ is the epicyclic frequency, $\Omega$ and $v_c =
\Omega/r $ are the angular rotation rate and circular velocity of the
unperturbed system at radius $r$, and $B \equiv -\kappa^2/4\Omega$.
The wavenumber of the spiral arm perturbation, $k$, is given by 
$k \equiv df(r)/dr$ 
where $\Phi_a(r) \propto e^{ i f(r)}$.  
$h_a$ corresponds to variations in the specific enthalpy of the same form as
(equation 1) and can be neglected in the limit $c_s << v_c$ which is
appropriate for the rotation curves of the low surface brightness
galaxies considered here.

Although these velocity perturbations become large near resonances,
a limit on the {\it minimum} possible velocity perturbation can be
placed from the observations.
We can rewrite the above equation as
\begin {equation}
v_{\phi a} = {ik \Phi_a  \over 
(1 - (m \Omega - \omega)^2 \kappa^{-2}) 2 \Omega}.
\end {equation}
Because the magnitude of the denominator of the above equation 
reaches a maximum near corotation we can place the approximate limit
\begin {equation}
\vert v_{\phi a} \vert \gta 
\left\vert  {k \Phi_a   \over  2 \Omega }\right\vert.
\end {equation}

A density perturbation 
of the same form as Eq.~3 causes the spiral potential
perturbation given above where
\begin {equation}
\Phi_a = {-2 \pi G \Sigma_a \over \vert k \vert}
\end {equation}
(\cite{B+T}, equation 6-17) for an infinitely thin disk.  Substituting this
into the above equation gives us the limit
\begin {equation}
\vert v_{\phi a} \vert \gta
\left\vert  { \pi G \Sigma_a  \over   \Omega } \right\vert.
\end {equation}
If we consider the possibility that the stellar component is more
massive than the gas component the above expression can be inverted to
yield an upper limit on the mass-to-light ratio, $M/L$, of the disk
given an upper limit on the $\phi$ component velocity perturbations
across the spiral arm:
\begin {equation}
M/L \lta {v_{\phi a} v_c \over \pi G S_a r}
\end {equation}
where $S_a$ is the surface brightness variation about radius $r$ (of
the same form as Eq.~3), and $v_{\phi a}$ is the maximum velocity
perturbation detected (or detectable in the case of no detection).

The $\phi$ component velocity perturbations are particularly
noticeable along the kinematic major axis where the line of sight
component of the velocity is in the azimuthal direction.  From Figures
1 and 2 we can see that for UGC~6614 these velocity perturbations are
smaller than $\sim 15$ km/s, whereas for F568-6 larger perturbations
are detected but are $\lta 25$ km/s.  Using these velocities as upper
limits and the $m=1$ and $m=2$ components of the surface brightness
variations from the azimuthal profiles (shown in Figure 5) upper
limits to the mass-to-light ratios as a function of radius for the two
galaxies are shown in Figure 7.

The $m=2$ component oval distortion for $r \sim 40''$ in UGC~6614 is
sufficiently strong to produce larger velocity perturbations than
detected if the mass-to-light ratio is greater than $\sim 3$ (for 
$R$ band in solar units) in this region.  From this we derive a limit $M/L
\lta 3$ in the central region of the galaxy.  Alternatively we can
think of this limit in terms of mass and say that if the disk has
perturbations of the same amplitude as observed in the $R$ band image,
the surface density of the disk must be less than $30 M_\odot/{\rm
pc}^2$ (for $M/L = 3$ and a surface brightness of 23.5 mag/arcsec$^2$
in the $R$ band at $r =$ 40\arcsec\ or 16 kpc).  For F568-6 the spiral
structure in the central regions is not as strong and only gives a
limit of $M/L \lta 6 $ at $r \sim$ 30\arcsec\ which is equivalent to a
surface density which must be less than $60 M_\odot/{\rm pc}^2$ (for
$M/L = 6$ and a surface brightness of 23.5 mag/arcsec$^2$ in the $R$
band at $r =$ 30\arcsec\ or 24 kpc).

The fits to the rotation curves of \cite{pic97} for both galaxies have
$M/L \lta 1$ for the bulge and disk and require a substantial dark
matter component even at small radii.  For these fits the peak of the
velocity contribution from the disk is only $\sim 40$ and 90 km/s for
UGC~6614 and F568-6 respectively.  For our upper limit of $M/L = 3 $
and 6 for the disks in UGC~6614 and F568-6 a substantial dark matter
component is still required to reach the observed maximum rotational
velocity.  Our upper limits reinforce the findings of \cite{pic97} and
others based on fits to the rotation curve that a substantial dark
matter component is required even in the central regions of these low
surface brightness galaxies.  This limit further requires that the
dark component cannot be in the form of a cold stellar disk (which
would have to be involved in the spiral structure).

The above high stellar surface densities are consistent with our
neglect of the gas density in Eq.~9 since they are significantly
higher than the gas density (see Figure 5).  However, even though
strong density variations variations exist in the \ion{H}{1} for these
galaxies we have estimated the minimum possible velocity response
using linear perturbation theory.  We note here that it is possible
that the non-linear response of the gas could cause the velocity
perturbations to be somewhat smaller than inferred from the above
limit.  In this case a higher mass-to-light ratio could be consistent
with the small size of the observed velocity perturbations.  If future
higher angular resolution \ion{H}{1} observations reveal larger
velocity perturbations along the spiral arms, then the upper limit for
the mass-to-light ratio would also be higher than stated here.
An underlying more massive disk at these radii could exist if it
had smaller azimuthal density variations than observed in the $R$ band.
For example if near-infrared observations (which might more
accurately trace surface density varitations) show smaller amplitude
variations in the spiral structure, $S_a$ in the above equation
would be smaller and a more massive disk could be consistent with the
observations.
The spiral arms observed in the $R$ band images show fine structure
which can only exist when the stellar disk is thin.
It is therefore reasonable to use the approximation of a thin disk 
for the potential (Eq.~7).

We also note that our lowest value for the mass-to-light ratio in UGC
6614 coincided with density perturbations that were part of an oval
distortion in the galaxy with position angle that varied only slowly
with radius.  While it is not unreasonable that this oval distortion
could be driving a strong gas response, the WKB or tight winding
approximation is no longer valid.  Although the minimum velocity
response is of the same order as given in Eq.~8, this equation
could be inaccurate by a factor of a few.


\section{Critical Forcing Required to Produce Shocks -- A Lower Limit for $M/L$}

The large density variations in the spiral structure of UGC~6614 and
F568-6 are evidence for shocks in the ISM induced by a spiral
gravitational potential.  Indeed the high arm/interarm density
contrast of \ion{H}{1} observed in galaxies such as M81 and M51 is one
of the major predictions of the spiral density wave theory.  In this
section we consider how much mass is required in the form of spiral
structure to drive shocks in the gas that would be consistent with the
$\sim$2:1 arm/interarm density contrasts observed in the \ion{H}{1} of
these two low surface brightness galaxies.

The response of the gas in a spiral density wave is primarily
dependent on the forcing gravitational field and the effective ISM
sound speed and only weakly dependent on the cloud dependent
properties such as the cloud mean-free path and the cloud number
density (\cite{rob84}, \cite{hau84}, and \cite{rob69}).  A critical
forcing parameter to produce shocks or large density contrasts in the
ISM was explored by \cite{rob69} and \cite{shu73}.  These authors
considered the role of $F$, the spiral gravitational force expressed
as percentage of the axisymmetric force.  Originally \cite{rob69}
found that a forcing of $F > 2\%$ was required to produce a density
contrast of greater than 2 and that for $F< 1\%$ no shock was
produced.  Subsequent work by \cite{shu73} found that the critical
forcing parameter depended on the effective sound speed of the gas,
$c_g$, and on the speed of the imposed spiral force $c_k\equiv
m\Omega_p/k$.  For a wide variety of pattern speeds, \cite{shu73}
found that a forcing amplitude of at least $3-4\%$ was required to
produce shocks in the gas, although when $c_k/c_g$ is moderately
greater than 1, a forcing of only $1\%$ could produce shocks.  This
differed from the critical forcing parameter introduced in
\cite{too77} (see also \cite{B+T}), valid in the limit of low
effective sound speed, which requires forcing amplitudes of a few
percent for moderate values of $c_k/c_g > 1$ (\cite{too77}).

Subsequent models and simulations of the gas response find that
forcing of at least a few percent is required to produce the observed
gas density contrasts.  For example using a forcing spiral
gravitational field based on photometry of M81 \cite{vis80} found that
when $F \lta 4\%$ no shocks were produced in his model.  Although
more recent modeling of M81 does not vary the forcing amplitude,
\cite{low94} produce the observed density contrast with simulations
driven by a spiral forcing amplitude of $5-10\%$.  By exploring the
properties of cloud-particle simulations \cite{rob84} found that
density contrasts of a few resulted where $F = 5-10\%$ using a spiral
structure based on M81 even for the low effective gas sound speed of
8 km/s.

To summarize, the critical forcing value of a few percent ($\gta$
3--4\%) seems to be required to produce shocks in the ISM that result
in significant gas density contrasts, although this value is dependent
on $c_g$ and $c_k$ so that an ISM with a low effective sound speed
forced by a fast spiral pattern could produce shocks with the
weak forcing amplitude of only $\sim 1\%$.

\subsection{Forcing Amplitudes in F568-6 and UGC~6614}

For the potential given in Eq.~3, the forcing amplitude
expressed as a ratio of the unperturbed axisymmetric gravitational
force can be written
\begin {equation}
F  \equiv \left\vert{\Phi_a k \over r \Omega^2}\right\vert.
\end {equation}
Using Eq.~7 we can rewrite this in terms of the density variation as
\begin {equation}
F  = {2 \pi G \Sigma_a  r \over v_c^2}.
\end {equation}

Using the above expression we have computed the forcing amplitudes
using $\Sigma_a$ estimated from the $m=1$ and $m=2$ gas and $R$ band
azimuthal components with $M/L =1$ in the $R$ band.
These forcing amplitudes are shown in Figure 8.  We note that the
forcing amplitudes are typically quite small.  The distortions in the
central regions of both galaxies are of sufficient strength to drive
shocks in the gas assuming a mass-to-light ratio of 1.  In these
regions the gas contribution to the spiral gravitational field are
negligible.  At larger radii the gas and stellar spiral gravitational
forcing in both galaxies are quite small, $\lta 2\%$ of the
axisymmetric force.

If we assume that a minimum particular forcing strength is required to
produce the observed \ion{H}{1} density contrasts, then we derive a
lower limit for the mass-to-light ratio of the disk.  We have chosen a
critical value of $2\%$ because even though a forcing of only $1\%$
might be able to cause shocks, is unlikely to cause the observed gas density
contrast.  Larger values typical of normal galaxies such as M81 might
be inconsistent with the low rate of star formation observed in these
galaxies (\cite{pic97}).  Assuming that forcing of at least $2 \%$ is
required to produce the observed \ion{H}{1} density contrasts then in
both galaxies we derive the lower limit $M/L \gta 1$ in the outer
parts of the disks ($r>60''$ or 24 kpc and $r>40''$ or 32 kpc in UGC
6614 and F568-6 respectively).  This can be expressed as a mass
density (if the underlying stellar disk has the same amplitude
azimuthal variations as observed in $R$ band).  In UGC~6614 for $r =$
60--120\arcsec\ the $R$ band surface brightness varies from 25--26
mag/arcsec$^2$ which for $M/L=1$ is equivalent to a disk stellar
surface density of 2.6--1.0 $M_\odot/{\rm pc}^2$.  In F568-6 for $r =$
40--90\arcsec\ the $R$ band surface brightness varies from 24--26
mag/arcsec$^2$ which for $M/L=1$ is equivalent to a disk stellar surface
density of 6.6--1.0 $M_\odot/{\rm pc}^2$.  These lower limits imply
that the stellar surface density is of the same order as the gas
surface density. This is consistent with the large scale morphology of
the spiral structure and the stability of the gas disk, both of which
suggest that a moderate stellar component is required to produce the
observed spiral structure.

We note that the forcing assuming $M/L=1$ given in Figure 8 (expressed
as a ratio of the unperturbed axisymmetric gravitational force) is
smaller than derived for normal high surface brightness galaxies but
not 10 times smaller as expected from the 1-3 lower magnitude surface
brightness disks of the low surface brightness galaxies, and the
equivalent size of their rotational velocites.  This is due to the
fact that at the large radii where spiral structure exists in these
low surface brightness galaxies, the axisymmetric force ($v_c^2/r$) is
lower than that at the smaller radii where the spiral structure exists in
normal galaxies.

One possible concern is that a moderate external tidal field could be
driving the gas response instead of a spiral disk component.  In this
case the disks of these galaxies could have very low mass-to-light
ratios.  However, UGC~6614 appears to be isolated; no gas near its
redshift was evident in the velocity channel maps of \cite{pic97}.  On
the other hand, in F568-6 there is evidence for interacting material;
there is a high velocity clump of gas to the south of the nucleus
which seems to be due to a superimposed, possibly interacting dwarf
(\cite{pic97}).
This scenario is unlikely, though, because an external tidal field
should cause an oval (non-spiral) perturbation to the gravitational
potential which would drive spiral shocks in the gas that are more
open than the tightly wound spiral arms observed.  It is also
difficult for an external tidal field to drive shocks in the gas over
a large range of radius.  An external tidal field could however be
ultimately responsible for exciting the spiral density waves in the
gas plus stellar disk (as is likely in the case of M51), but then the
resulting stellar and gas derived spiral gravitational field must be
sufficiently strong to cause shocks in the gas as assumed here.

\section{Summary and Discussion}

The spiral structure of the low surface brightness galaxies F568-6 and
UGC~6614 is large scale, with arms that wrap more than half a
revolution, and extend out to 50 and 80 kpc in UGC~6614 and F568-6
respectively.  These spiral arms are visible in the $R$ band images,
the \ion{H}{1} column density maps and as kinks in the \ion{H}{1} velocity
fields.  The density contrasts observed in the \ion{H}{1} maps are
high, with arm/interarm contrasts of $\sim 2:1$, whereas the velocity
perturbations due to spiral structure are low, in the range 10-20 km/s
and 10-30 km/s in UGC~6614 and F568-6 respectively.

We use the small velocity response to place upper limits on the
mass-to-light ratio of the stellar disk.  The strongest limits occur
at small radii ($r \sim 40''$ or 16 kpc and $r \sim 30''$ or 24 kpc in
UGC~6614 and F568-6 respectively) where there are strong distortions
observed in the $R$ band images.  The weak observed response in the
$\phi$ velocity component limits the mass-to-light ratios of the disk
in these regions to $M/L < 3$ and 6 for UGC~6614 for F568-6
respectively.  This is equivalent to requiring the densities of the
disks (if they have azimuthal variations of the same size as that
observed in the $R$ band images) to be less than 30 and 60
$M_\odot/{\rm pc}^2$ at a radius of 16 and 24 kpc for UGC~6614 and
F568-6 respectively.  An underlying more massive disk at these radii
could exist if it had smaller azimuthal density variations than observed in
the $R$ band.  These limits are sufficiently strong to require a
significant dark matter component even in the central regions of this
galaxy, confirming the findings of previous studies.  Our limits
furthermore imply that this dark matter component cannot be in the
form of a cold disk since a cold disk would necessarily be involved in
the spiral structure, though a hot disk cannot be excluded.  We note
that this upper limit was derived assuming a linear gas response in
the tight winding or WKB approximation, and that a non-linear gas
response driven by bar like or oval distortions could cause small
velocity perturbations to be present even in the case of stronger
potential perturbations (or higher mass-to-light ratios).

To produce the large arm/interarm \ion{H}{1} density variations it is
likely that the spiral arm potential perturbation is sufficiently
strong to produce shocks in the gas.  For a forcing that is greater
than $2\%$ of the axisymmetric force, $M/L \gta 1$ is required in both
galaxies in the outer regions.  This is equivalent to a disk surface
density between $r =$ 60--120\arcsec\ in UGC~6614 of 2.6--1.0
$M_\odot/{\rm pc}^2$ and between $r =$ 40--90\arcsec\ in F568-6 of
6.6--1.0 $M_\odot/{\rm pc}^2$ assuming that the amplitude of the
variations in the disk mass is the same as that observed in the $R$
band.  These lower limits imply that the stellar surface density is of
the same order as the gas surface density.  This is consistent with
the large scale morphology of the spiral structure, and the stability
of the gas disk, both of which suggest that a moderate stellar
component is required to produce the observed spiral structure.  
The gas disks alone probably fall below the critical gas
density (\cite{pic97}) emperically found by \cite{ken89} 
for the onset of masive star formation
and so are not likely to be unstable enough to support spiral density waves. 
However it is likely that the combined stellar and gas
disks are (see \cite{jog84} for instability in a two fluid disk).  The
coupled gas and stellar disk would be consistent with the large scale
spiral arm morphologies which do not resemble that of gas dominated
flocculent galaxies.

The limits for the mass-to-light ratio of the disk derived here
suggest that the disks of these two low surface brightness galaxies
lie in the range $1 < M/L < 6$ (in the $R$ band).  This range is
identical to that found for normal higher surface brightness disk
galaxies derived from maximal disk fits to the observed rotation
curves (\cite{ken87a}, \cite{ken87b}).  In other words the disk
mass-to-light ratio limits placed here are not abnormal compared to
those of normal galaxies.  Our limits on the disk surface densities
remain consistent with previous studies which find that low surface
brightness galaxies have substantially lower mass surface densities
than normal galaxies (\cite{deb96b}, \cite{spray95b}).

In low surface brightness galaxies 
since the disk contributions to the rotation curves
are small compared to the halo, good fits to the rotation curves 
can be acheived with a range of mass-to-light ratios.  
Fits to the rotation curves using halo profiles
of the form proposed by \cite{nav95} yeilded the best fit mass-to-light
ratios of $\sim 0.8$ and $0.5$ for UGC~6614 and F568-6 respectively
(\cite{pic97}),   whereas
\cite{imp97} using an isothermal halo found a good fit
to the rotation curve of F568-6 with a much higher $M/L_B = 8$. 
The disk mass-to-light ratios derived from these fits probably
depend upon the halo profile assumed and 
whether the bulge and disk are allowed to have different mass-to-light
ratios and so are not tightly constrained.

We note that multi-wavelength observations (particularly those in the
infrared) find that amplitude variations across spiral arms can be a
strong function of wavelength (e. g. \cite{rix93}).  While the low
\ion{H}{1} column depths in UGC~6614 and F568-6 suggest that
extinction from dust is not a large effect in these low surface
brightness galaxies, it would not be surprising if an older population
of stars (which would be more apparent in the near infrared
wavelengths) might have smaller spiral arm amplitudes.  In this case a
lower limit for the mass-to-light derived as we have done here but
from a near infrared image might yield even stronger limits requiring
an even more massive stellar disk.  If high mass-to-light ratios are
required, then a mass-to-light ratio variation across the disk could be
required to yield a good fit to the rotation curve assuming a smooth
halo component, and a large dark matter component might not
necessarily be required in the central regions.  Multi-wavelength
observations coupled with metallicity measurements should also provide
useful information about what kinds of stellar populations would be
consistent with the range of mass-to-light ratios given here.

It might prove to be interesting to place similar mass-to-light ratio
limits in low surface brightness dwarf galaxies which also can have
spiral structure (for example UGC~11820, UGC~5716, and F568-1, from
\cite{vz97} and \cite{deb96b}).  These galaxies can have regular,
symmetrical \ion{H}{1} velocity fields and smooth \ion{H}{1} column density 
maps, even when the optical components lack symmetry or contain strong
spiral structure.  The optical components therefore don't strongly
influence the velocity field (although some of these effects might be
visible in \ion{H}{1} observations at higher angular resolution).
This suggests that the upper limits discussed here might be
particularly revealing.  

Strong spiral structure detected in \ion{H}{1} in the 
outer regions of higher surface brightness galaxies could 
be used to estimate the mass density of a low surface
brightness stellar disk.  
Some interesting candidates for such a study might be 
the dark blue compact dwarf NGC~2915 which may
contain spiral structure well outside its optical disk (\cite{meu96}),
and the polar ring galaxy NGC~4650A which shows evidence for 
spiral arms in its ring (\cite{arn97}).

\acknowledgments

We acknowledge helpful discussions and correspondence with 
R. Kennicutt,  L. van Zee,  A. Nelson, G. Rieke, J. Navarro and H-W. Rix.
We acknowledge support from NSF grant AST-9529190 to M. and G. Rieke.

\clearpage


\newpage
\clearpage
 
\begin{figure*}
\caption[junk]{ a) $R$ band image shown as grayscale with the
\ion{H}{1} velocity contours.  The contour spacing is 20 km/s with the
lowest contour at 6180 km/s. The left side is the approaching side.
Some of the background exponential disk has been subtracted from the $R$
band image to make the spiral structure more apparent.  Note that the
spiral arms are at locations of kinks in the velocity field.  
b) \ion{H}{1} intensity (grayscale) with \ion{H}{1} velocity
contours. The beamsize for the \ion{H}{1} observations is plotted in
the upper right corner.  Note that the surface density contrast is
high with an arm/interarm contrast of $\sim$2:1.  The high gas
density contrast is a consequence of shocks in the ISM and one of the
predictions of strong spiral density waves.  In contrast to gas-rich
flocculent galaxies, the morphology of the spiral structure is large
scale extending more than half a revolution.  This suggests that a
stellar component, gravitationally coupled to the gas, is involved in
the spiral density waves.
\label{fig1} }
\end{figure*}

\begin{figure*}
\caption[junk]{ a) $R$ band image of F568-6 shown as grayscale with
the \ion{H}{1} velocity contours.  The contour spacing is 20 km/s with
the lowest contour at 13600 km/s. The right side is the approaching
side.  Some of the background exponential disk has been subtracted
from the $R$ band image to make the spiral structure more apparent.
Note that the spiral arms are at locations of kinks in the velocity
field.
b) \ion{H}{1} intensity (grayscale) with \ion{H}{1} velocity
contours. The beamsize for the \ion{H}{1} observations is plotted in
the upper right corner.  Note that the surface density contrast is
high with an arm/interarm contrast of $\sim$2:1.  The high gas density
contrast is a consequence of shocks in the ISM and one of the
predictions of strong spiral density waves.  In contrast to gas-rich
flocculent galaxies, the morphology of the spiral structure is large
scale extending more than half a revolution.  This suggests that a
stellar component, gravitationally coupled to the gas, is involved in
the spiral density waves.
\label{fig2} }
\end{figure*}

\begin{figure*}
\caption[junk]{ Azimuthal profiles as a function of azimuthal angle of
the deprojected UGC~6614 $R$ band image (solid line) given in
mag/arcsec$^2$ (left vertical axes) and the deprojected \ion{H}{1}
surface density image (dotted line) in $M_\odot/{\rm pc}^2$ (right
vertical axes).  On the horizontal axis, $0^\circ$ is along the major
axis of the galaxy (PA $=287^\circ$), and $90^\circ$ is perpendicular
to this axis in the plane of the galaxy which lies along PA
$=17^\circ$.  The radius of each azimuthal cut is given on the left
hand side of the plot.  For small radii ($35''<r<55''$) there is an
oval distortion causing the large density variations seen in the
azimuthal profiles within these radii.  At larger radii, brighter
regions lie exclusively along the faint outer spiral arms.  The
north-eastern spiral arm is brighter than the south-western one.
Although peaks in the \ion{H}{1} and $R$ band surface brightness are
correlated (see Figure 1) they do not correspond to maximum densities
or surface brightnesses in the azimuthal cuts except at large radii.
Better correlation might be observed in higher resolution \ion{H}{1}
data.  At larger radii, \ion{H}{1} dense regions lie exclusively along
the outer spiral arms.
\label{fig3} }
\end{figure*}

\begin{figure*}
\caption[junk]{ Azimuthal profiles as a function of azimuthal angle of
the deprojected F568-6 $R$ band image (solid line) given in
mag/arcsec$^2$ (left vertical axes) and the \ion{H}{1} surface density
image (dotted line) in $M_\odot/{\rm pc}^2$ (right vertical axes).  On
the horizontal axis, $0^\circ$ is along the major axis of the galaxy
(PA $=-105^\circ$), and $90^\circ$ is perpendicular to this axis in
the plane of the galaxy which lies along PA $=-15^\circ$.  The radius
of each azimuthal cut is given on the left hand side of the plot.  For
small radii ($20''<r<45''$) there is an oval distortion causing the
large density variations seen in the azimuthal profiles within these
radii.  Although peaks in the \ion{H}{1} and $R$ band surface
brightness are correlated (see Figure 2) they do not correspond to
maximum densities or surface brightnesses in the azimuthal cuts.
Better correlation might be observed in higher resolution \ion{H}{1}
data.
\label{fig4} }
\end{figure*}

\begin{figure*}
\caption[junk]{ 
a) Azimuthally averaged, inclination corrected $R$
band surface brightness profile for UGC~6614.
b) The $m=2$ (solid points) and $m=1$ (open points) component
azimuthal variations in $R$ band for UGC~6614 expressed as a percentage
of the azimuthally average value.  There is a strong oval distortion
at  $r\sim 40''$.  Bi-symmetric or two arm spiral structure is apparent
as strong $m=2$ components at large radii in UGC~6614.
c) Azimuthally averaged, inclination corrected $R$ band surface
brightness profile for F568-6.  
d) The $m=2$ (solid points) and $m=1$ (open points) component
azimuthal variations in $R$ band for F568-6 expressed as a percentage of
the azimuthally average value.  There is a moderate distortion at 
$r\sim 30''$.
\label{fig5} }
\end{figure*}

\begin{figure*}
\caption[junk]{ 
a) Azimuthally averaged, inclination corrected gas
density profile derived from the \ion{H}{1} column density map of UGC~6614.
b) The $m=2$ (solid points) and $m=1$ (open points) component
azimuthal variations in the gas for UGC~6614 expressed as a percentage
of the azimuthally average value.
c) Azimuthally averaged, inclination corrected gas density profile
derived from the \ion{H}{1} column density map of F568-6.  
d) The $m=2$ (solid points) and $m=1$ (open points) component
azimuthal variations in the gas for F568-6 expressed as a percentage
of the azimuthally average value.  Contrasts are higher at larger
radii in both galaxies.
\label{fig6} }
\end{figure*}

\begin{figure*}
\caption[junk]{ 
a) Upper limits for the mass-to-light ratio, $M/L$, (in
solar units) in UGC~6614 based upon the $R$ band $m=2$ (solid points)
and $m=1$ (open points) component azimuthal variations assuming that
tangential velocity perturbations are no greater than 15 km/s.
b) Upper limits for the mass-to-light ratio, $M/L$, (in solar units) in
F568-6 based upon the $R$ band $m=2$ (solid points) and $m=1$ (open
points) component azimuthal variations assuming that tangential
velocity perturbations are no greater than 25 km/s.
\label{fig7} }
\end{figure*}

\begin{figure*}
\caption[junk]{ 
a) Spiral forcing in UGC~6614 expressed as a
percentage of the axisymmetric force caused by the $R$ band surface
brightness variation (hexagons) and the \ion{H}{1} surface density
(triangles) for the $m=2$ (solid points) and $m=1$ (open points)
component variations assuming a mass-to-light ratio of $M/L =1$ in
solar units in $R$ band.  Note that the spiral forcing is typically
quite small.
b) Same as a) but in F568-6.
\label{fig8} }
\end{figure*}


\clearpage
\clearpage
\newpage

\begin{deluxetable}{lrrrrr}

\footnotesize
\scriptsize
\tablewidth{300pt}
\tablecaption{Observation Parameters and Derived Properties\tablenotemark{a}}
\tablehead{
\colhead{} & 
\colhead{F568-6} & 
\colhead{UGC~6614}  
}
\startdata
Total absolute R band magnitude           & $-23.6\pm0.1$ & $-22.3\pm 0.1$ \nl
Distance assumed (Mpc)                    & $184$         & $85$           \nl
Total      HI    line flux (Jy km/s)      & $4.4\pm 0.3$  & $15.0\pm 0.8$  \nl
Total      HI    mass ($10^{10} M_\odot$) & $3.6\pm 0.4$  & $2.5\pm 0.2$   \nl
FWHM of synthesized beam ($'' \times ''$)  & $19.5 \times 18.4$ &  $21.4 \times 19.8$ \nl
P.A. of synthesized beam (degrees)      & $-73.0$ & $52.3$  \nl
Channel spacing (km/s)                  & $22.6$  & $10.7$  \nl
RMS noise in channel maps (mJy/beam)    & $0.25$  & $0.7$   \nl
Limiting column density,\tablenotemark{b} ~$N_H$ ($10^{19} {\rm cm}^{-2}$) & $4.4$  & $5.8$   \nl
\enddata
\tablenotetext{a}{From \cite{pic97}} 
\tablenotetext{b}{Three sigma dectection in one channel} 
\end{deluxetable}

\end{document}